Impurity impact ionization avalanche in p-type diamond

V. Mortet[1-3] and A. Soltani[4]

[1] CNRS ; LAAS; 7 Avenue du Colonel Roche, F-31077 Toulouse, France.

[2] Université de Toulouse; UPS, INSA, INP, ISAE, UT1, UTM, LAAS, F-31077 Toulouse, France

[3] IMO, Hasselt University, Wetenschapspark 1, 3590 Diepenbeek, Belgium

[4] IEMN/CNRS 8520, Cité scientifique, Avenue Poincaré, Université Lille Nord de France, 59652 Villeneuve d'Ascq, France

Abstract

Electrical conductivity of a highly boron doped chemical vapor deposited diamond thin film has been studied at different temperatures under high electric field conditions. Current-voltage characteristics have been measured using pulsed technique to reduce thermal effects. Experimental results evidence deep impurity impact ionization avalanche in p-type diamond up to room temperature.

The investigation of hot carriers in semiconductor, i.e. carriers subjected to high electric field, began in the mid 1950's. Early work began on n-type Germanium with the study at liquid helium temperature of impact ionization of shallow donors, i.e. with activation energy ($E_a$) about 10 meV. At these low temperatures, Germanium has very few carriers and it is very resistive because of the insufficient thermal energy to ionize impurities. Under a critical electrical field, the few stray carriers present in Germanium acquire sufficient energy to ionize impurities to produce new carriers by inelastic collisions. This effect leads to an increase of several orders of magnitude in carrier density until the complete ionization of impurities in an avalanche process. This phenomenon is called *impurities* impact ionization avalanche (IIIA) and it has to be distinguished from the destructive *lattice* impact ionization avalanche causing the formation of electron holes pairs that occurs at higher field strengths. Several applications of IIIA had been shown[1-3], including terahertz electroluminescence in silicon[4,5]. However these applications are limited to cryogenic temperature as IIIA is only observed at temperature where donor or acceptor impurities are frozen out, i.e. cryogenic temperature in most semiconductors[6-10].

At the first glance, Diamond appears to be the best semiconductor material for high power and high frequencies electronic applications because of its outstanding properties such as electrons and holes high mobilities, high breakdown field and high thermal conductivity. However, the high activation energy of electrically active impurities: nitrogen ($E_a$=1.7 eV), phosphorous ($E_a$=0.6 eV) and boron ($E_a$=0.37 eV)[11], compared to thermal energy at room temperature (kT=25 meV), in another words, the lack of low ionization energy impurities burdens diamond as semiconductor for electronic applications. On another hand, the high activation energy of diamond's doping impurities is of a particular interest for devices based on impurity impact ionization avalanche operating at room temperature and above because only a small amount of impurities are ionized in non degenerated doped diamond. We report in this letter the experimental evidence of Deep Impurity Impact Ionization Avalanche (DIIIA) in boron doped diamond up to room temperature.

The boron doped diamond layer studied in this work was grown on Ib (100) oriented single crystal diamond substrates by Plasma Enhanced Chemical Vapor Deposition[12]. It was grown in a 1% mixture of methane diluted in hydrogen and a B/C ratio of 500 ppm using trimethilboron. The thickness of the boron doped layer has been estimated to be *2 μm* from double mass measurement. After deposition, the sample was cleaned in hot sulfuric acid mixed with potassium nitrate, rinsed in hot DI water and dried with nitrogen.

A pair of tungsten electrodes *10 μm* wide with a gap of 10μm were deposited by sputtering and patterned by lift-off technique on the sample to make test resistors. Tungsten contacts were used instead of Ti/Al because aluminum is quickly damage by electro-migration during measurements. Measurements were made in the dark, in an Oxford Instruments cryostat and in Helium at atmospheric pressure. The temperature in the cryostat was varied between 175K and 300K. Current-voltage characteristics were measured using pulsed measurement technique to prevent thermal effects on the measurements[8]. Single pulses (0-5V) produced by an Agilent 81101A pulse generator are amplified up to 500V pulses using a homemade amplifier (×100) based on a high voltage power operational amplifier PA89 from Cirrus Logic. These pulses are applied to the diamond resistor in series with a 1 kΩ resistance. The pulse's amplitude and the voltage drop in the series resistance are measured using a Tektronix TDS 3012 oscilloscope to determine the current-voltage characteristic of the doped diamond resistors.

Fig. 1 shows the typical time dependence of the voltage *V(t)* and the current *I(t)* across a diamond resistor at a temperature of 175 K. The rising time of the pulses is severely limited by the amplifier's characteristics. Nevertheless, one can clearly observe the nonlinear and sudden rise of the current at voltage of *V ~ 450 V,* i.e. an electric field of *450 kV.cm$^{-1}$*. Because of the current flat at the voltage plateau is stable, we assume that there is no significant thermal effect in the measurements. An increase of the temperature in the test device structure would lead to an increase of carriers' concentration and current at the voltage plateau. We can also rule out that the current rise is due to electric breakdown in the helium atmosphere, as the measured breakdown voltage is bellow the breakdown voltage of helium[13]. Furthermore, the measured current remains limited and there is no voltage drop as in a gas breakdown. We assume that the current rise is due to a carrier multiplication. Fig. 2 shows the current voltage characteristic at different temperatures from 175 K up to 300 K in logarithmic and linear scales. Each point of Fig. 2a were measured at the plateau of the pulses, whereas the I(V) curve of fig. 2b was plotted from the time dependent current and voltage curves. Three different regions can be observed in these current-voltage characteristics. Bellow 100 V bias, the current follows the Ohm's law, i.e. the drift velocity of the carrier increases linearly with the electric field. One can see on Fig. 2a the regular increase of the current with temperature in the ohmic region. This normal increase is due the increase of holes density by thermal activation. The activation energy $E_a \sim 0.28\ eV$ has been determined from an Arrhenius plots using the variation of the resistance of the sample as a function of the temperature. This activation energy is lower than Boron activation energy (0.37 eV) and it is significant of a highly boron doped (about $\sim 2\text{-}5\times10^{19}$ cm$^{-3}$) diamond layer[14,15] in a temperature range of incomplete ionization with compensation[11]. At higher bias, a non linear increase of the current, that can be call pre-breakdown region, is attributed to carriers' multiplication. A further increase in the voltage leads to a sudden and sharp increase of the current like in an avalanche breakdown phenomenon. It is interesting to observe that for the same current level, for instance 3 mA, the semiconducting diamond is either in the breakdown region or in the pre-breakdown region depending on the temperature. This result is consistent with the breakdown in diamond rather than in the gas. The non linear increase of current cannot be due to an increase of the carrier drift velocity because of the large increase of current but it must be due to an increase of the carrier density. Multiplication of the carrier density is attributed to the deep impurities impact ionization of boron acceptor by the few and energetic holes under the influence of the large electric field between the two 10 µm spaced electrodes. The energy of the hot carriers must be higher than the ionization energy of boron in diamond to trigger the DIIIA. For comparison, the calculated mean hole energy in a electric field of 100 kV.cm$^{-1}$ in silicon determined by Monte-Carlo technique is about 0.4 eV[16]. Because of the power capability of our experimental set-up the current in the diamond resistor is limited and the total ionization of boron impurities cannot be observed. Using a transmission line pulse measurement setup[17], a currents as high as 3 A have been measured at room temperature.

That corresponds to current density of $j \sim 10^7\ A.cm^2$ that is consistent with the level of current density of the sample with all impurities ionized. The breakdown field ($E_{brk}$), i.e. the highest field at which the current rise vertically, varies from ~350 kV.cm$^{-1}$ to ~ 450 kV.cm$^{-1}$ as the temperature decreases from room temperature to 175 K. Fig. 3 shows the variation of the ratio (β) between the measured current ($I_m$) and the ohmic, i.e. linear, extrapolation of the ohmic current ($I_\Omega$) in the high voltage/field region as a function of the voltage in a semi-log graph. An example of the linear extrapolation is shown on fig. 4a for a temperature of 225 K. This ratio is representative of the carrier's multiplication between the electrodes. This figure shows the carrier multiplication rises with the field and its decrease with the temperature. The rise of carrier multiplication is higher and stepper at low temperature. Several effects influence the carrier multiplication by impurities impact ionization: the impurity concentration, the ionized impurities concentration and the temperature. Temperature rise decreases the carriers' multiplication coefficient because it increases phonon's scattering and decreases the energy gained by the carriers in the electric field[16]. As a result, the carrier multiplication occurs at larger voltage as the temperature increases (see fig. 4 and 6).

In summary, I-V characteristics of a resistor made on a highly doped CVD diamond layer measured at high electric field in pulsed mode at different temperature are reported. The observed nonlinear and sudden current rise of current at high electric filed has been attributed to boron impurity impact ionization avalanche in the sample. These results are the experimental evidences of deep impurity impact ionization avalanche in p-type diamond. This effect is observed up to room temperature because of the high activation energy of boron impurities in diamond. The multiple measurements made during this study proves that the observed breakdown in is non destructive and reproducible.

[16] M. Levinshtein, J. Kostamovaara, S. Vainshtein, *Breakdown phenomena in semiconductors and semiconductor devices* (World Scientific, Selected Topics in Electronics and Systems – vol. 36, 2005).

[17] Ph.D thesis, J. J. Ruan, Universite Toulouse III – Paul Sabatier, France, 2010.

Figures caption

Fig. 1: Waveform of the pulsed voltage and pulsed current across the diamond resistor at T = 175K.

Fig. 2: Current-voltage characteristic measured by pulsed technique in log-log scale (a) and linear scale (b) from 300 K to 175 K temperature range.

Fig. 3: Variation of the ratio ($\beta$) between the measured current ($I_m$) and the ohmic, i.e. linear, extrapolation of the current ($I_\Omega$) in the high voltage/field region as a function of the voltage

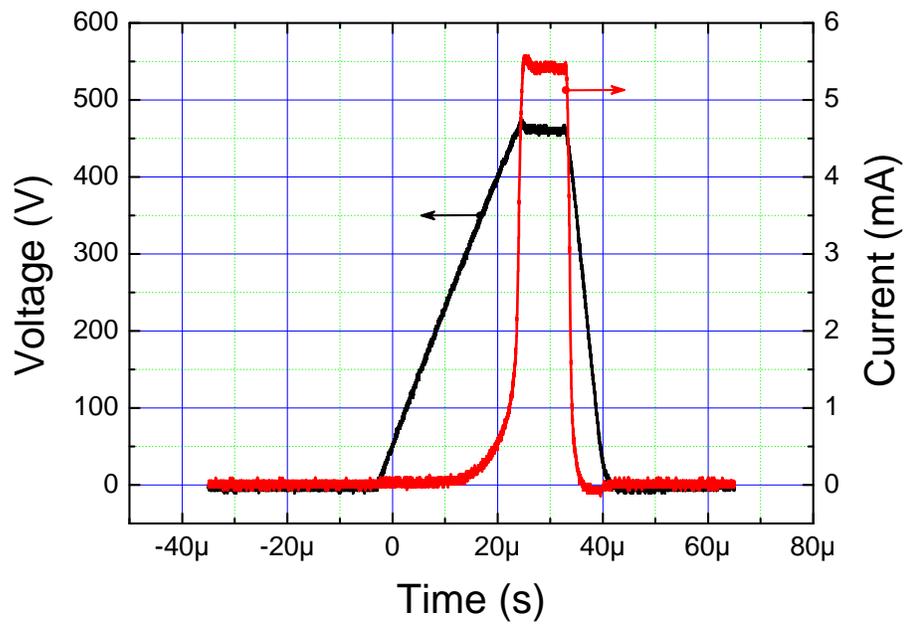

Fig. 1

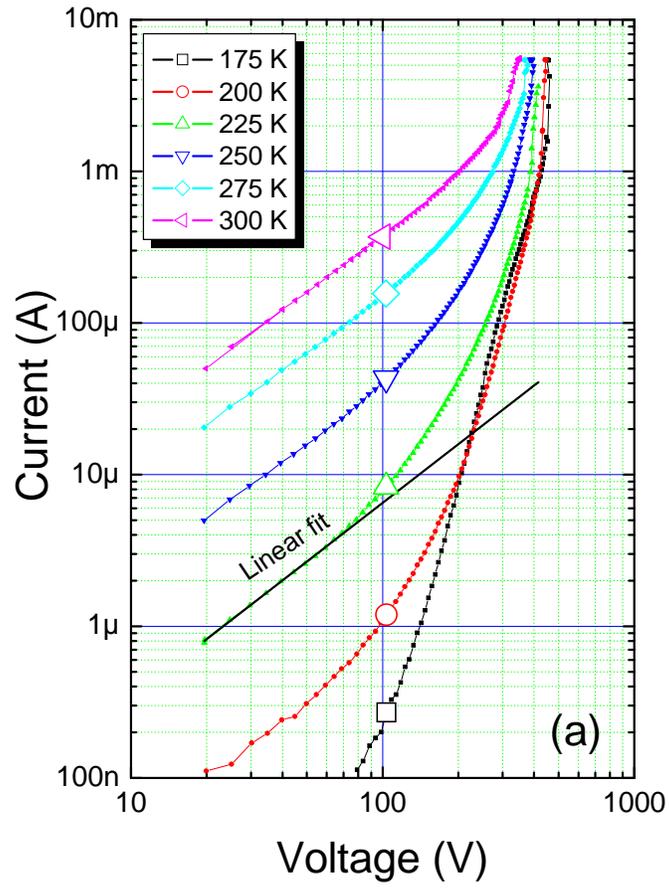

Fig. 2a

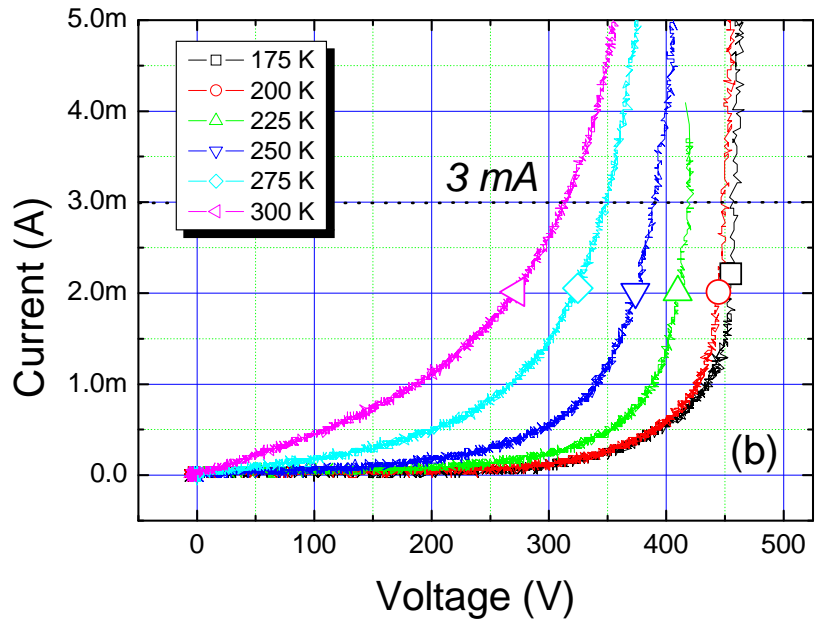

Fig. 2b

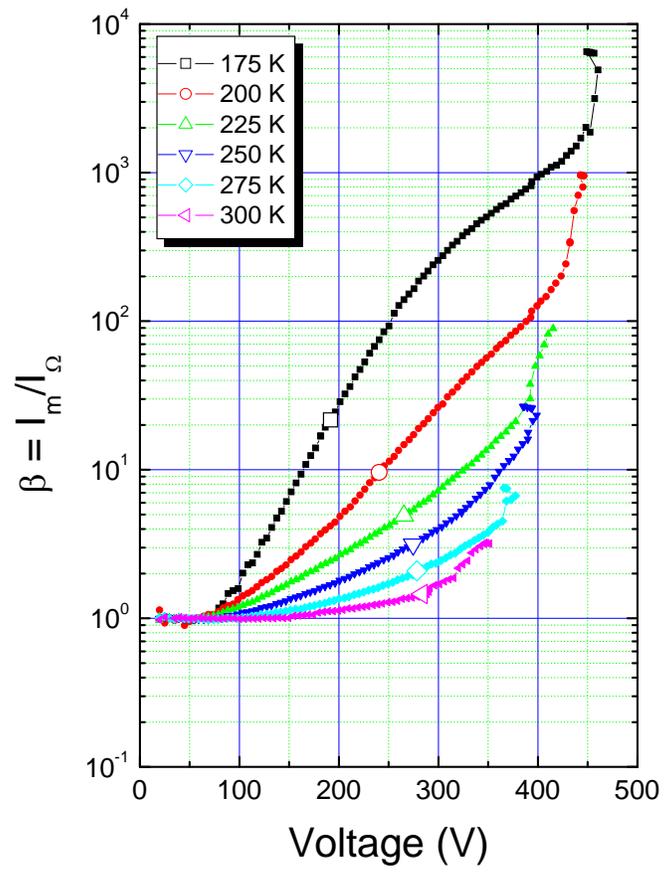

Fig. 3